\documentclass[12pt]{iopart}

\usepackage{amssymb,graphicx,amsbsy,hyperref,color,txfonts}

\begin{document}

\title[Community consistency determines the stability transition window]{Community consistency determines the stability transition window of power-grid nodes}
\author{Heetae Kim$^1$, Sang Hoon Lee$^{1,2}$, and Petter Holme$^{1,3,4}$
}
\address{$^1$Department of Energy Science and $^2$Integrated Energy Center for Fostering Global Creative Researcher (Brain Korea 21 plus), Sungkyunkwan University, Suwon, 440--746, South Korea \\
$^3$IceLab, Department of Physics, Ume{\aa} University, Ume{\aa}, 90187, Sweden \\
$^4$Institute for Future Studies, Stockholm, 10131, Sweden}
\ead{holme@skku.edu}

\begin{abstract}
The synchrony of electric power systems is important in order to maintain stable electricity supply. Recently, the measure \textit{basin stability} was introduced to quantify a node's ability to recover its synchronization when perturbed. In this work, we focus on how basin stability depends on the coupling strength between nodes. We use the Chilean power grid as a case study. In general, basin stability goes from zero to one as coupling strength increases. However, this transition does not happen at the same value for different nodes. By understanding the transition for individual nodes, we can further characterize their role in the power-transmission dynamics. We find that nodes with an exceptionally large transition window also have a low community consistency. In other words, they are hard to classify to one community when applying a community detection algorithm. This also gives an efficient way to identify nodes with a long transition window (which is computationally time consuming). Finally, to corroborate these results, we present a stylized example network with prescribed community structures that captures the mentioned characteristics of basin stability transition and recreates our observations.
\end{abstract}



\maketitle

\section{Introduction}
\label{sec:introduction}

Electric power systems are often modeled as oscillators on networks~\cite{Kuramoto1984,Acebron2005,Arenas2008}. The rotational motion of power generators for alternating currents needs to be synchronized with a rated frequency of an electric power grid~\cite{Filatrella2008,Motter2013,Menck2013,Menck2014,Schultz2014,Nishikawa2015}. Stable synchrony of electric power grids' nodes is important to prevent cascading failures. When a node gets an external perturbation leading the node away from a synchronous state, then the node either returns to the synchronous state or escapes from its basin of attraction, typically to a different limit cycle. Basin stability is calculated as the fraction of the possible phase values a node can be perturbed to and still recover synchronicity~\cite{Menck2013}. It has become the standard way to quantify a power-grid node's robustness to large point perturbations. The basin stability is determined by the magnitude of the perturbation and network characteristics of the node. It also depends on the coupling strength between the nodes in the dynamic system representing the electric power flow. For instance, Refs.~\cite{Menck2014,Schultz2014} investigate topological characteristics of particularly stable or unstable nodes. The authors conclude that, in general, dead ends weaken the basin stability, while detours strengthen it. Such simple indirect estimates are helpful because basin stability is very computer intensive to calculate.
 
These previous studies, however, study basin stability for one specific coupling strength~\cite{Menck2013,Menck2014,Schultz2014}. However, basin stability is sensitive to the coupling strength. In a power grid, the transmission capacity, which determines to the coupling strength, is decided based on the physical conditions such as the air temperature, the length of transmission line, and the amount of transmitting electricity. When the amount of electricity transmitted exceeds the transmission capacity, it may cause the breakdown of a large part of the system. Therefore, it is important to understand the functional dependence of basin stability on the coupling strength, not only its behavior at a fixed and intermediate coupling strength. We use the real electric power grid in Chile as our prime example, but also discuss how the results generalize.
To be more specific, we investigate the network-structural factors~\cite{Barrat2008,Newman2010} behind a node's basin stability as a function of the coupling strength. Based on the Chilean case study, we argue that there is a connection between the coupling strength dependence and community structure---the way the network can be decomposed into communities that are densely connected within and sparsely connected between each other~\cite{Porter2009,Fortunato2009}. Other authors have noticed that those communities easily synchronize internally. In a similar spirit, we investigate how the membership strength of a node within a community decomposition (its community consistency) of the network relates to the coupling strength dependence, specifically the width in parameter space of the basin stability's transition from zero to one.

\section{Methods}
\label{sec:methods}

\subsection{The dynamical model of electricity transmission}
\label{sec:dynamical_model}

The synchronization dynamics between power-grid nodes is commonly modeled as the set of nonlinear oscillators using a Kuramoto-type model~\cite{Kuramoto1984,Acebron2005,Filatrella2008,Motter2013,Menck2013,Menck2014,Schultz2014,Nishikawa2015}:
\begin{equation}
\ddot \theta_i = \dot \omega_i = P_i - \alpha_i \dot \theta_i - K \sum_{i \ne j} a_{ij} \sin(\theta_i - \theta_j) ,
\label{eq:Kuramoto_type_equation}
\end{equation}
where $P_i$ is the net power generation (positive) or consumption (negative) at node $i$; $\alpha_i$ is a dissipation constant; $K$ is the coupling strength between nodes $i$ and $j$; the adjacency matrix $a_{ij} = 1$ if there is an edge between node $i$ and $j$, and $a_{ij} = 0$ otherwise; $\theta_i$ is the phase of node $i$; $\omega_i = \dot \theta_i$ is the angular velocity of node $i$. A power grid is considered to be stable and synchronized when $\omega_i$ vanishes for all of the nodes so that the system maintains the desirable constant frequency. For numerical integration, we use the Runge-Kutta method~\cite{NR} with the convergence criterion~\cite{Menck2014} of the time derivative of angular frequency $\dot\omega < 5\times10^{-2}$ and the angular frequency $\omega < 5\times10^{-2}$ based on the actual fluctuations, i.e., we consider the initial phase and angular velocity $(\theta_i,\omega_i)$ of $i$ as belonging to the stable basin if the system eventually converges to $\dot\omega < 5\times10^{-2}$ and $\omega < 5\times10^{-2}$ after the $i$ is perturbed. We perturb $i$ by initializing $\omega_i$ and $\theta_i$ to random values in the intervals $[-100,100]$ and $[-\pi,\pi]$. The other nodes have their values of the synchronous state. Following Refs.~\cite{Menck2013,Menck2014,Schultz2014}, we set $\alpha_i = 0.1$ for all $i$.

\subsection{The Chilean power grid}
\label{sec:Chilean_power-grid}

We construct the Chilean power grid, our prime example, from the operational records published by the National Energy Committee of Chile~\cite{CNE2012} and the data regarding the power-grid transmission lines published by Centro de Despacho Econ{\'o}mico de Carga del Sistema Interconectado Central (CDEC-SIC)---the major electric power company in Chile~\cite{CDEC-SIC}. The Chilean power grid consists of power plants and substations as nodes, and the transmission lines between them as edges~\cite{Barrat2008,Newman2010}. A power plant generates electricity and a substation distributes it to the final consumers. We aggregate the power grid using Zhukov's aggregation scheme~\cite{Machowski2008}, resulting in nodes being either net generators ($P_i > 0$ in Eq.~(\ref{eq:Kuramoto_type_equation})) or net consumers ($P_i < 0$) based on the real data~\cite{CDEC-SIC}. After determining the sign of $P_i$ values, we first set $P_i = -1$ for $P_i < 0$ and $P_i = P_\mathrm{pos} > 0$ so that $\sum_i P_i = 0$ (Kirchhoff's circuit laws). The inactive nodes such as towers or net zero traders ($P_i = 0$ in Eq.~(\ref{eq:Kuramoto_type_equation})) do not intervene with the perturbation propagation and are removed. 
 We apply the so-called Kron reduction~\cite{Kron1939,Dorfler2013} procedure. It is a node elimination technique widely used in electrical circuit design. For a given network configuration, Kron reduction removes a node and adds new edges between neighbors of the eliminated node at each step~\cite{Kennelly1899,Rosen1924}. Consequently, as the more nodes are Kron-reduced, the denser (in terms of mean degree) the network becomes. However, even though the network structure is changed, the new edges are  placed so that the network maintains the relationship between nodes in terms of electricity flow. For the Chilean power grid, we eliminate 46 inactive nodes using Kron reduction. As a result, the final network consists of 291 net consumers, 129 net generators, and 573 edges. 

\subsection{Basin stability transition}
\label{sec:Basin_stability_transition}

Basin stability is a stability measure for a network~\cite{Menck2013,Menck2014,Schultz2014}. For each node, an initial phase value is selected uniformly at random. Then we run the dynamics until convergence (according to the criterion presented in Sec.~\ref{sec:dynamical_model}). The fraction of initial phase values that leads to convergence defines the basin stability. We sample 100 points uniformly at random from the intervals $-100 < \omega < 100$ and $-\pi < \theta < \pi$. Assuming a uniformly random sampling is done for the sake of simplicity (following Refs.~\cite{Menck2013,Menck2014,Schultz2014}). Note that the basin stability of a node could be compared to other nodes in the same network, but not necessarily to nodes in other networks.

For the so-called infinite bus-bar model (basically a mean-field approximation of the problem we study), the basin stability increases gradually with the transmission capacity and suddenly becomes unity (stable)~\cite{Menck2013}. This jump appears when the transmission capacity is large enough for all perturbations or initial shocks from the entire phase space to be absorbed into the system. However, in the more realistic multi-node model, the basin stability and its transition form are different for different nodes. So for a complete picture, every node needs to be analyzed separately.

\begin{figure}
\hfill\includegraphics[width=0.85\textwidth]{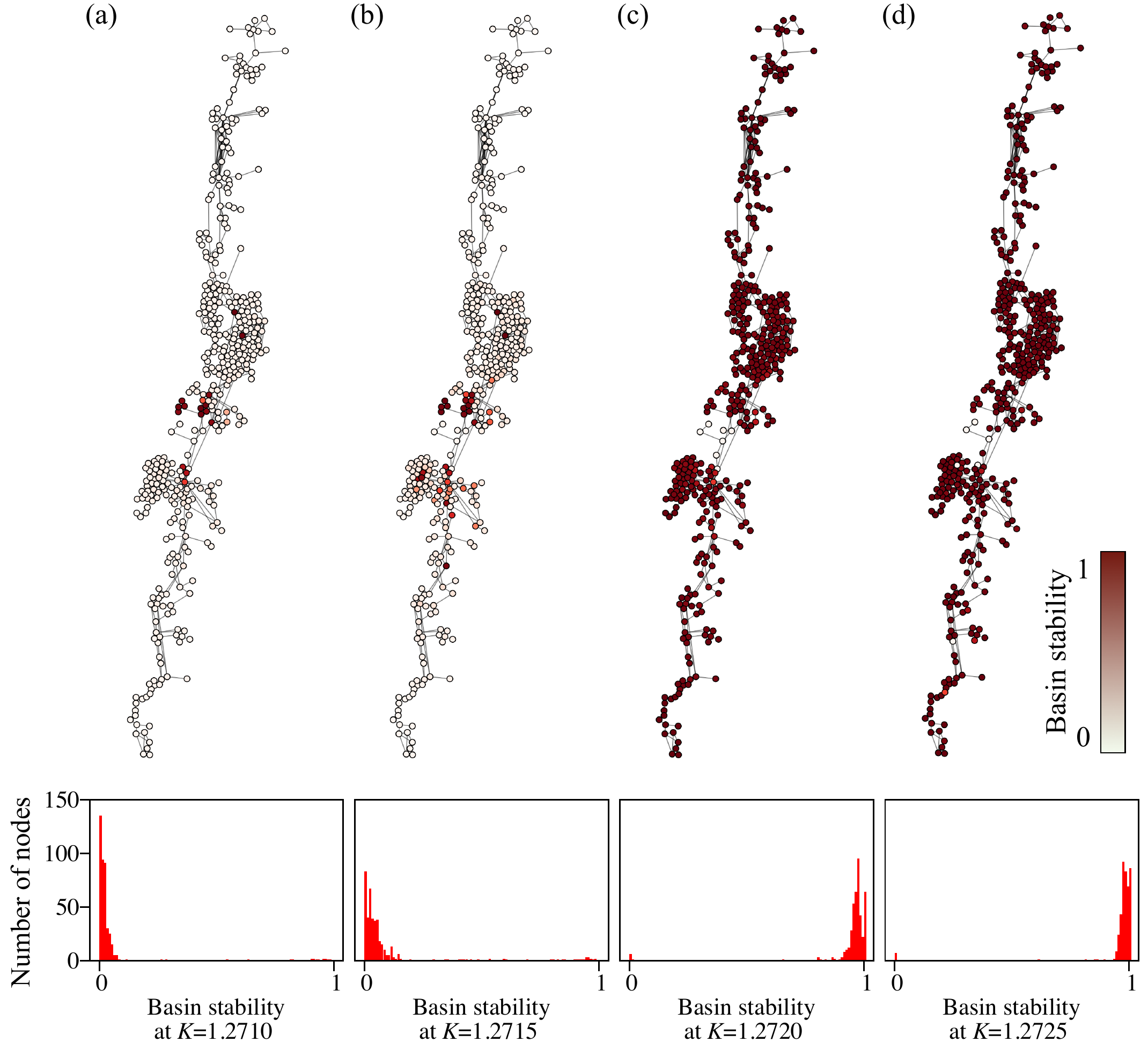}
\caption{Color-map and distributions of basin stability values of nodes in the Chilean power grid for different coupling strength $K = 1.2710$ (a), $1.2715$ (b), $1.2720$ (c), and $1.2725$ (d). 
\label{fig:basin_stability_for_different_coupling_strength}}
\end{figure}

We show the results for the distribution of basin stabilities of the Chilean power grid's nodes as functions of the coupling strength $K$ in fig.~\ref{fig:basin_stability_for_different_coupling_strength}.  For low $K$, all nodes have zero basin stability. As $K$ increases, a few nodes start to reach a non-zero basin stability (fig.~\ref{fig:basin_stability_for_different_coupling_strength}(a)). When the coupling strength reaches a certain threshold, almost all the nodes suddenly turn into a stable state. Note that the basin stability shows a nonlinear transition as a function of coupling strength. The dramatically different distribution of basin stabilities for the different $K$ reinforces the importance of considering not only a certain basin stability at a $K$ value but also the transition of basin stability corresponding to different $K$ values. In Sec.~\ref{sec:results}, we address the transition of basin stability and specifically introduce basin stability transition window.

\section{Results}
\label{sec:results}

\subsection{Basin stability transition window}
\label{sec:basin_stability_transition_window}

\begin{figure}
\hfill\includegraphics[width=0.85\textwidth]{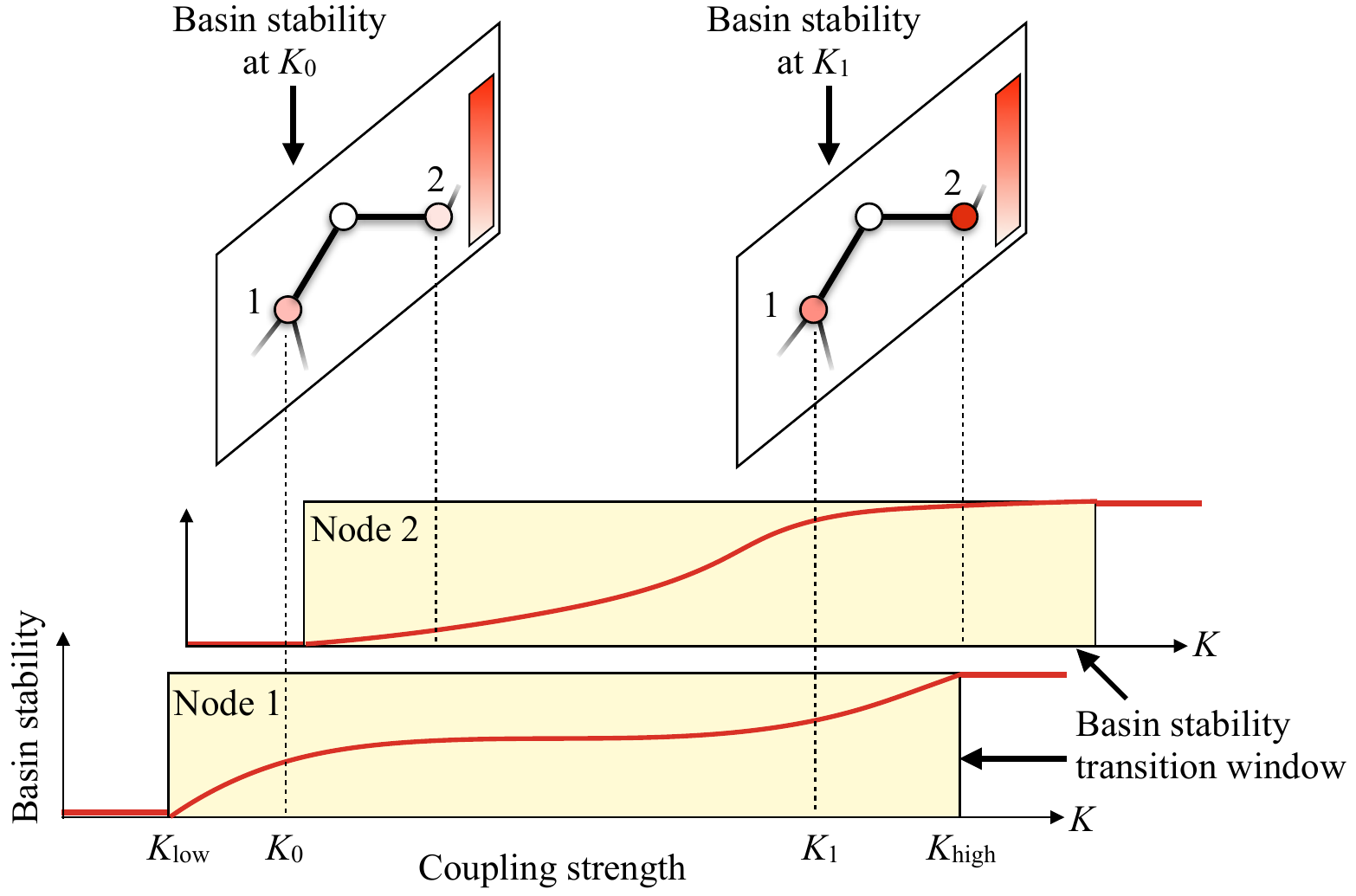}
\caption{A schematic diagram of basin stability transition as a function of coupling strength on a power grid. 
}
\label{fig:schematic_diagram}
\end{figure}

Basin stability changes, as mentioned, with the coupling strength. Figure~\ref{fig:schematic_diagram} shows a schematic diagram of the basin stability transition pattern that we call \emph{transition window}, for two nodes of a hypothetical network. At $K = K_0$, node $1$ has larger basin stability than node $2$. On the other hand, when the coupling strength increases to $K = K_1$, node $2$ becomes more stable than node $1$. These transitions of basin stability of nodes $1$ and $2$ are represented as a function of $K$ on the transition window (the shaded rectangles of fig.~\ref{fig:schematic_diagram}).

To locate the transition windows of the nodes, we use the bisection method~\cite{Burden1985}. We start by finding the lower bound of the transition window $K_\mathrm{low}$, which is the minimum value of $K$ that makes the basin stability larger than zero. We do this by tracking the basin stability from the $K_\mathrm{min} = K_\mathrm{min,init}$ and $K_\mathrm{max} = K_\mathrm{max,init}$, and measure the basin stability of a node at the midpoint $K_\mathrm{midpoint} = (K_\mathrm{min} + K_\mathrm{max}) /2$. When the basin stability at $K_\mathrm{midpoint}$ is non-zero, $K_\mathrm{max}$ is set to the current $K_\mathrm{midpoint}$ value and the new $K_\mathrm{midpoint}$ value is recalculated from the new $K_\mathrm{max}$ value. Similarly, when the basin stability at $K_\mathrm{midpoint}$ is zero, the new $K_\mathrm{min}$ is set to the current $K_\mathrm{midpoint}$. This iteration continues until $K_\mathrm{max} - K_\mathrm{min} < K_\mathrm{threshold}$ and $K_\mathrm{low}$ is set to the final values of $(K_\mathrm{min} + K_\mathrm{max})/2$. The upper bound $K_\mathrm{high}$ of the transition window (the maximum value of $K$ that makes the basin stability smaller than unity) is also determined in the same fashion. As a result, we obtain the transition width $\Delta K = K_\mathrm{high} - K_\mathrm{low}$ that represents the width of the transition window. We use $K_\mathrm{min,init} = 0$, $K_\mathrm{max,init} = 20$, and $K_\mathrm{threshold} = 0.01$.

\begin{figure}
\hfill\includegraphics[width=0.85\textwidth]{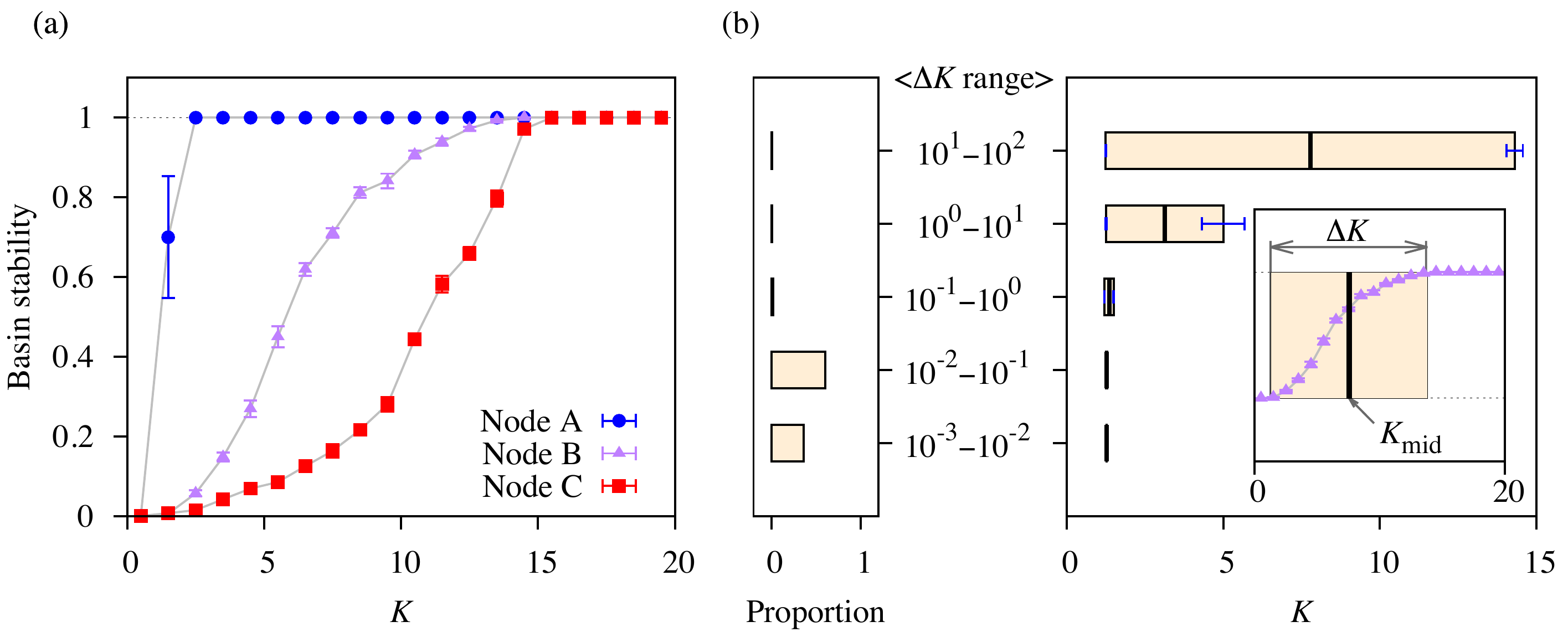}
\caption{Basin stability transition of node A, B, and C (see fig.~\ref{fig:model_network_result} for the actual locations of the nodes) (a) and average transition windows (b). In (a), the solid gray lines are the guide to the eyes. The dots represent the average basin stability of every ten points with the resolution $0.1$ from $K = 0$ to $19.9$ (each dot located at $\overline{K}$ corresponds to the averaged values of basin stability for $\overline{K} - 0.5$, $\overline{K} - 0.4$, \dots , $\overline{K} + 0.4$), where error bars corresponding to the standard deviation for each of the uniform bins with the size $1.0$. Nodes A, B, and C reach the complete basin stability (i.e.\ unity) at $K = 1.3$, $13.1$, and $14.7$, respectively. In fig.~\ref{fig:delta_K_distribution}(b), on the left panel, the fraction of nodes belonging to each logarithmic bin is shown for each bin. On the right panel, the average $\Delta K$ and $K_\mathrm{mid}$ are shown for each bin, where the error bars show the standard deviation values of $K_\mathrm{low}$ (on the left) and $K_\mathrm{high}$ (on the right). The inset shows the $\Delta K$ and $K_\mathrm{mid}$ values in a transition window (the filled rectangle) for node B.}
\label{fig:delta_K_distribution}
\end{figure}

$\Delta K$ can vary much between the nodes. For instance, node A (one of the nodes with the narrowest $\Delta K$ values) of the Chilean power grid shown in fig.~\ref{fig:delta_K_distribution}(a) undergoes the sudden basin stability jump at $K \simeq 1.3$ and reaches the completely stable state at a smaller $K$ value than other nodes such as nodes B and C as comparison. The distribution of $\Delta K$ in the Chilean power grid is highly right-skewed (with a mean of $0.154$ and median of $0.022$). Figure~\ref{fig:delta_K_distribution}(b) shows various statistics of the transition window, where we divide the $\Delta K$ values into five different logarithmic bins and present the fraction of nodes belonging to each bin (the left panel of fig.~\ref{fig:delta_K_distribution}(b)) and the range of average $[K_\mathrm{low}, K_\mathrm{high}]$ for each bin (the right panel of fig.~\ref{fig:delta_K_distribution}(b)). The vast majority (about 96\%) of nodes have very small values of $\Delta K < 0.1$ (about 35\% of nodes even have $\Delta K < 0.01$). However, there are a few nodes with relatively very large $\Delta K$ values, e.g., the average range $[K_\mathrm{low}, K_\mathrm{high}]$ for the bin corresponding to the maximum $K$ range ($10^1$--$10^2$) is more than 13 times as large as that for the bin corresponding to the minimum $K$ range ($10^{-3}$--$10^{-2}$). In general, the wider of the transition window a node is, the larger $K$ value in the middle ($(K_\mathrm{low} + K_\mathrm{high})/2$, henceforth denoted as $K_\mathrm{mid}$) the node has.

It is interesting to note that the transition curves themselves seem to have different shapes for different nodes, even for the nodes with similar $\Delta K$ values. For example, $\Delta K$ of nodes B and C have the similar $\Delta K$ values but the former (latter) node shows roughly concave (convex) curves at the transition. The different range of $\Delta K$ and the shape can be interpreted in several ways. If a node has small $\Delta K$, the location of $K_\mathrm{mid}$ is also small according to our observations, so the synchrony of the node tends to be stable even for small values of coupling strength. In other words, the node stays functional and is hardly affected by the accidental performance drop of the transmission line connected to it. However, at the same time, once the basin stability value started to decrease, the node suddenly loses the stability. Consequently, it makes the detection of system failure difficult. When the transition of the basin stability is gradual, on the other hand, a system disturbance can be detected by a drop in the basin stability. In summary, narrow transition windows could be good for keeping stability, while the wider window are better for an early detection of system failure.

\subsection{Chilean power-grid analysis}
\label{sec:community_consistency}

One drawback of basin stability as a robustness measure of nodes in power-grids, is that it is computationally demanding. Therefore, it is desirable to find less computationally restrictive, indirect measures to estimate the basin stability. Some previous studies proposed such topological indicators for basin stability---Menck \emph{et al.}~\cite{Menck2014} predict that dead-end nodes have small basin stability, while Schultz \emph{et al.}~\cite{Schultz2014} predict that detour nodes have large basin stability. 

The basin stability estimators in Refs.~\cite{Menck2014} and \cite{Schultz2014} assume a specific $K$-value. If $K$ is not precisely known or varying, it can only identify extreme cases (very stable or unstable nodes). The width of transition window is thus a more appropriate measure to capture the role of a node in the transmission dynamics. To understand why some nodes have larger $\Delta K$ than others, we try to find the most appropriate explanatory measure. This is far from trivial. Indeed phase synchronization on a network is a complex consequence of the propagation of phase difference and recovery due to the interactions between nodes~\cite{Arenas2008}.

It is known that the mesoscopic properties of networks, such as community structure, play a significant role in the synchronization on networks~\cite{Arenas2006,TZhou2007,EOh2008}. A community is a subnetwork, that is more strongly connected within than to the rest of the network~\cite{Porter2009,Fortunato2009}. Different community detection methods could divide the same network in different ways, but there would typically be pairs of nodes that always are classified as belonging to the same community. Some community detection methods are non-deterministic and could also give a different community decomposition between different runs of the same algorithm. One could therefore define nodes often classified to the same community as having high \emph{community consistency} with respect to this algorithm.
We use GenLouvain~\cite{GenLouvain}, a variant of conventional Louvain community detection algorithm~\cite{Blondel2008} known for its computational scalability, to identify communities in networks. We can also control the resolution parameter in the modularity function used as the objective function to maximize, to detect communities with different scales~\cite{Fortunato2007,Good2010,Lancichinetti2010}. We define community consistency of individual nodes as the nodes' degree of certainty of the assigned community. Previously, Refs.~\cite{HKwak2011,Lancinchinetti2012} calculate community consistency (or ``consensus clustering'') among assigned communities in multiple runs of each stochastic community detection algorithm, where they focus on the community consistency of each algorithm. In this paper, however, we measure community consistency of \emph{individual nodes} to relate it to the nodes' basin stability transition width $\Delta K$.

\begin{figure}
\hfill\includegraphics[width=0.85\textwidth]{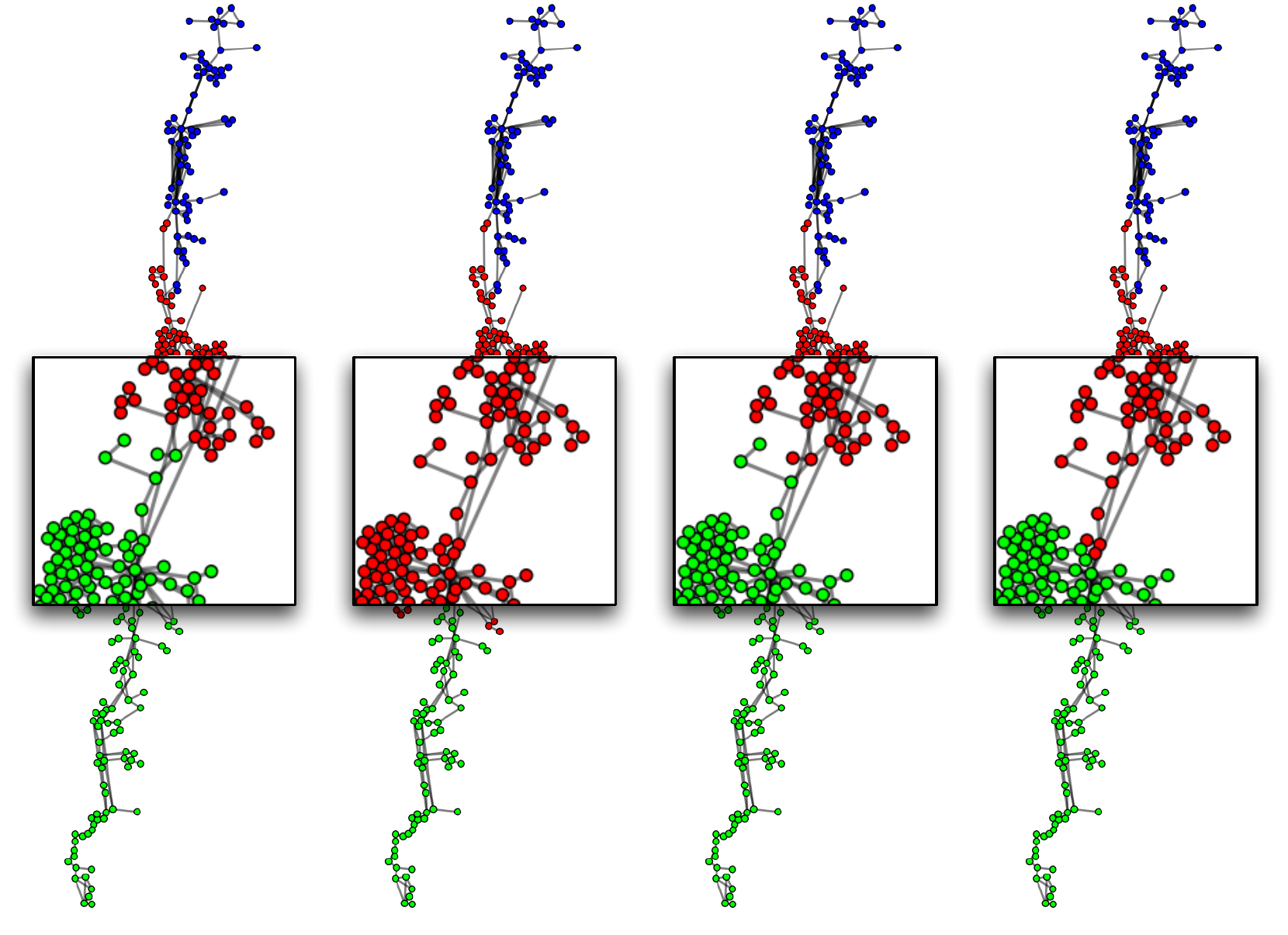}
\caption{Four different community detection results are shown for the same resolution parameter $\gamma = 0.0323$, where the regions of nodes with inconsistent community assignments are magnified. Different colors represent different communities detected by the GenLouvain method.
}
\label{fig:result_magnified}
\end{figure}

Figure~\ref{fig:result_magnified} shows four different community detection results of the GenLouvain method on the Chilean power grid, where the we use the same resolution parameter~\cite{Fortunato2007,Good2010,Lancichinetti2010}---$\gamma = 0.0323$. We chose this value to get around three communities which matches our visual impression of the network. As hinted in fig.~\ref{fig:result_magnified}, most nodes are assigned to the same community for all iterations. We call such nodes \textit{community consistent}. Some nodes, however, are assigned to different communities in different runs (the magnified region in fig.~\ref{fig:result_magnified}). In terms of power-grid dynamics, such nodes could, be believe, be influenced by perturbations from different directions (communities).

To measure the community consistency of individual nodes, we first observe a node $i$ is perfectly community consistent if, and only if, it gets always classified into a cluster with the same other nodes. Let $\phi_{ij}$ be the fractions of runs of the community detection algorithm when $i$ and $j$ are classified into the same community, then $i$ is perfectly community consistent if and only if $\phi_{ij}$ is either $1$ or $0$ for all $j$. Since $\phi_{ij}\in[0,1]$, we can obtain a metrics for community consistency by measuring the average distance to $\phi_{ij}=1/2$. But rather than the linear distance ($|\phi_{ij}-1/2|$), we sum the square distance. First, this conforms our measure to standard measures of deviation or spread like the root-mean-square, variance, and radius of gyration. Second, the non-linearity of the parabola accentuates values close to $0$ or $1$ and tones down the mid-interval values. This turns out to be practical since most $\phi$-values are close to $0$ or $1$. Finally, we multiply a factor $4$ to the outlined measure to get a value in the unit interval. In summary, the community consistency of $i$ is given by
\begin{equation}
\Phi_i = \frac{1}{N-1}\sum_{j\neq i} (1-2\phi_{ij})^2.
\label{eq:community_consistency}
\end{equation}

\begin{figure}
\begin{center}
\includegraphics[width=0.6\textwidth]{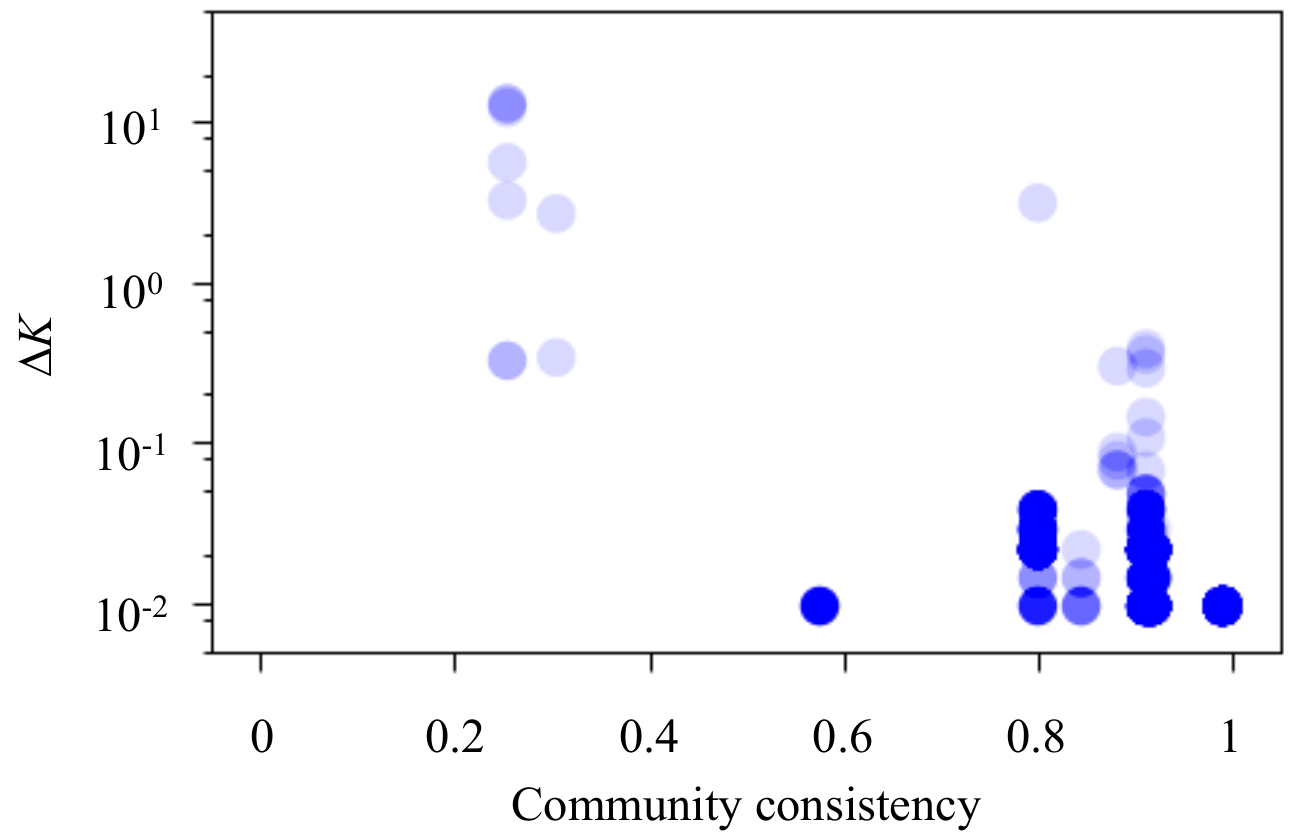}
\end{center}
\caption{$\Delta K$ and community consistency. This is basically a scatter plot but darker points represent more points, drawn with transparency. 
}
\label{fig:delta_K_vs_CC}
\end{figure}

\begin{table}
\caption{Pearson correlation coefficient $r$ of $\Delta K$ versus community consistency ($\Phi$), degree ($k$), clustering coefficient ($C$), and current flow betweenness ($F$) centrality. 
}
\begin{center}
\begin{tabular}{l|cccc}
\hline
 & $\Phi$ & $k$ & $C$ & $F$ \\
\hline
$r$ & $-0.581$ & $0.033$ & $-0.054$ & $0.072$ \\
$p$-value & $< 10^{-3}$ & $0.500$ & $0.266$ & $0.139$ \\
\hline
\end{tabular}
\end{center}
\label{table:correlation}
\end{table}

\begin{figure}
\hfill\includegraphics[width=0.85\textwidth]{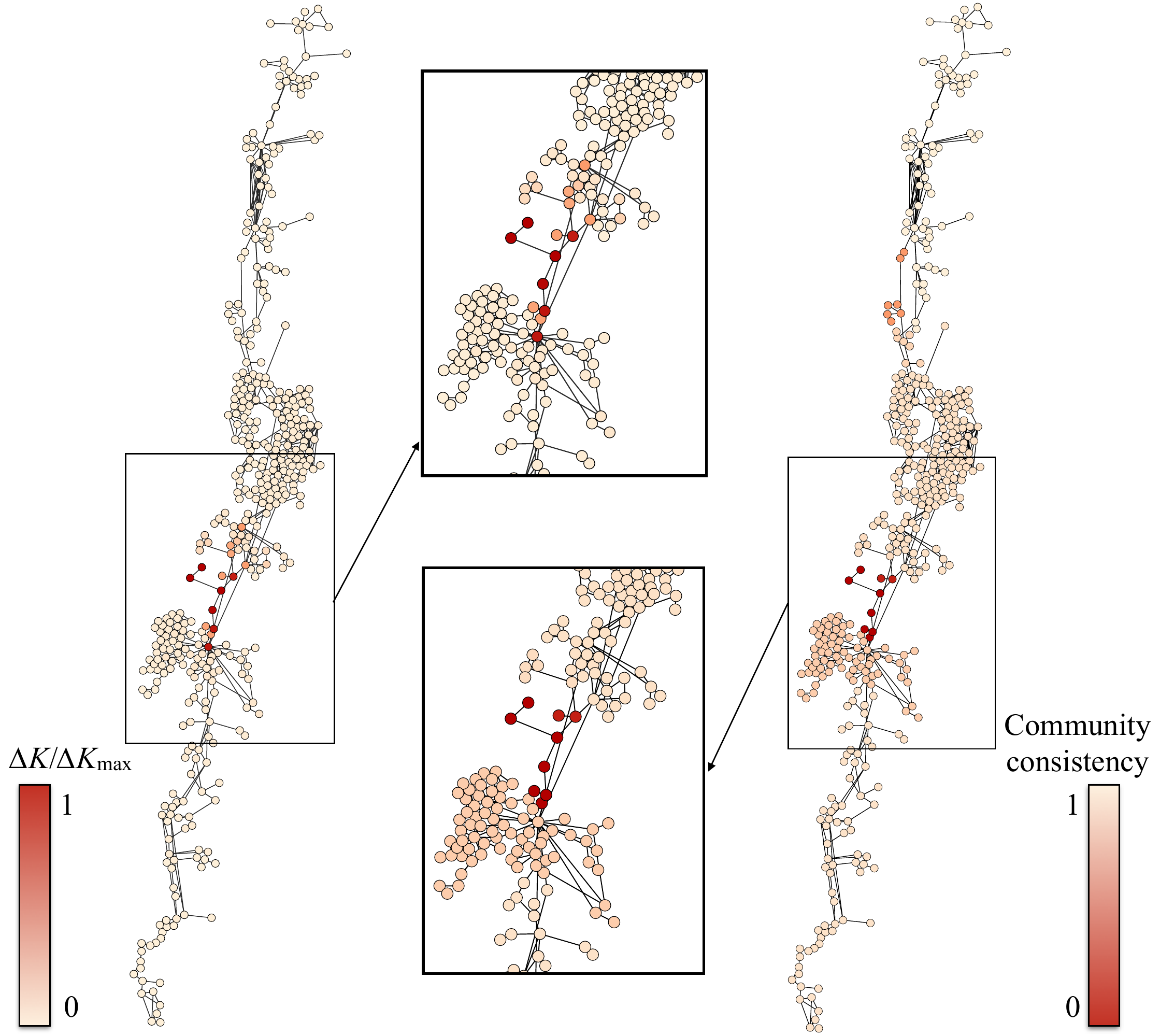}
\caption{$\Delta K / \Delta K_\mathrm{max}$ (the left panel) and community consistency (the right panel) of the Chilean power grid, where $\Delta K_\mathrm{max}$ is the maximum $\Delta K$ value among the nodes. The insets show the area of interest where the nodes with large $\Delta K$ and small community consistency.
}
\label{fig:delta_K_and_CC_on_the_map}
\end{figure}

The correlation between community consistency and $\Delta K$ of basin stability transition window of nodes is shown in fig.~\ref{fig:delta_K_vs_CC} and Table~\ref{table:correlation}, where the larger values of community consistency a node has, the narrower $\Delta K$ the node has. In other words, if a node's community consistency is weak, so that the node is assigned to different communities for each different realization, the node tends to have a wide basin stability transition window. 
Figure~\ref{fig:delta_K_and_CC_on_the_map} illustrates the nodes where $\Delta K$ and community consistency is colored, which provides a visual evidence of our result of the correlation between $\Delta K$ and community consistency. It seems that the basin stability transition occurs at some characteristic values of coupling strength for different communities of nodes as a unit, and the transition threshold is somehow smeared out for those nodes with weak or inconsistent community membership, which makes $\Delta K$ large. There are several nodes in the northern part of the power grid with lower community consistency, but we believe that the effect of the nodes is not strong enough to affect basin stability transition window. Moreover, compared to the actual numerical integration for calculating the basin stability, the  calculating community consistency is much faster (at least for reasonably fast  community detection algorithms).

\begin{figure}
\hfill\includegraphics[width=0.85\textwidth]{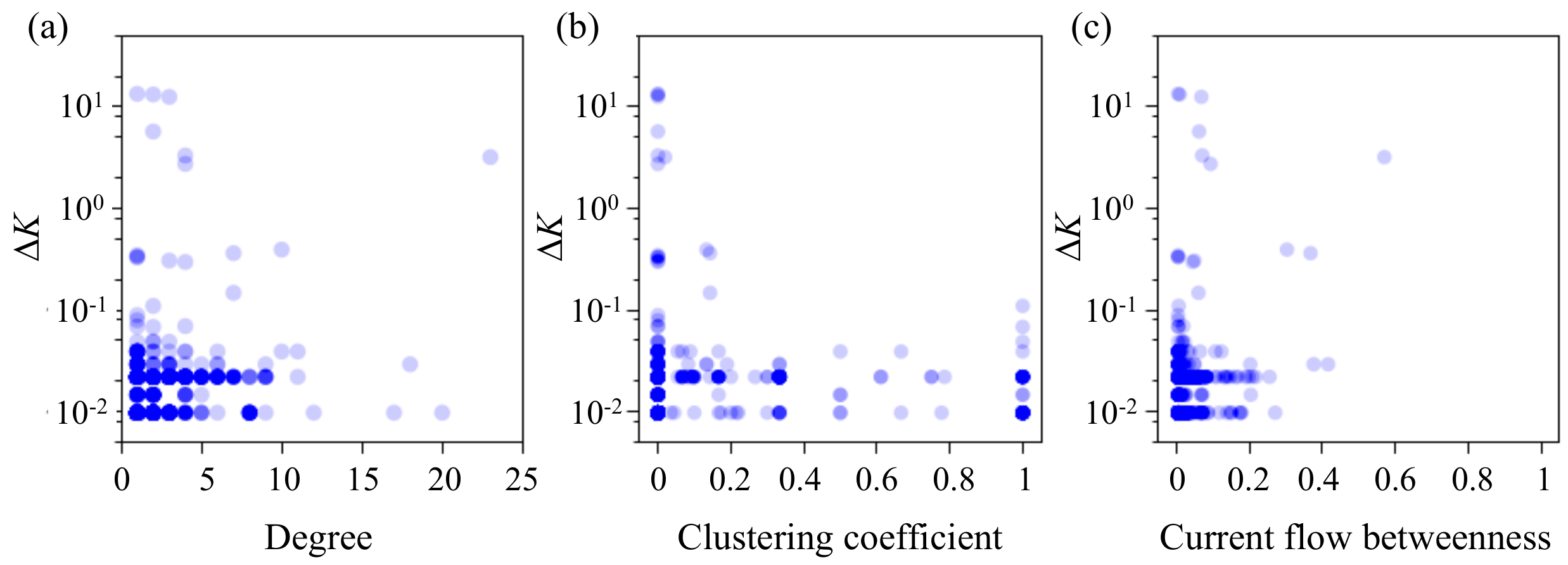}
\caption{Correlation between $\Delta K$ and degree (a), clustering coefficient (b), and current flow betweenness (c) centralities. These are basically scatter plots but darker points represent more points, drawn with transparency. 
}
\label{fig:correlation_with_centralities}
\end{figure}

To check if there is any other (possibly simpler) network measures that can predict $\Delta K$, we measure other network metrics. The Pearson correlation coefficients for $\Delta K$ versus degree~\cite{Barrat2008,Newman2010} and versus community consistency are about $0.033$ ($p$-value around $0.5$) and $-0.581$ ($p$-value less than $10^{-3}$), respectively. For the correlation coefficient values for other representative network centralities and community consistency are shown in Table~\ref{table:correlation}, indicating that only the mesoscopic measure of community consistency is significantly correlated with $\Delta K$, in contrast to other conventional measures such as degree, clustering coefficient~\cite{Newman2010}, and current flow betweenness centralities (supposedly a more relevant type of betweenness in our power-grid case than the ordinary one)~\cite{Newman2010}. The two former centralities are microscopic or local, while the latter is macroscopic as it deals with all of the possible pathways for an entire network. 
We tried other measures than the ones presented in Table~\ref{table:correlation}, but we could not find anything statistically significantly correlated with $\Delta K$.

\subsection{Example network analysis}
\label{sec:model_network}

\begin{figure}
\hfill\includegraphics[width=0.95\textwidth]{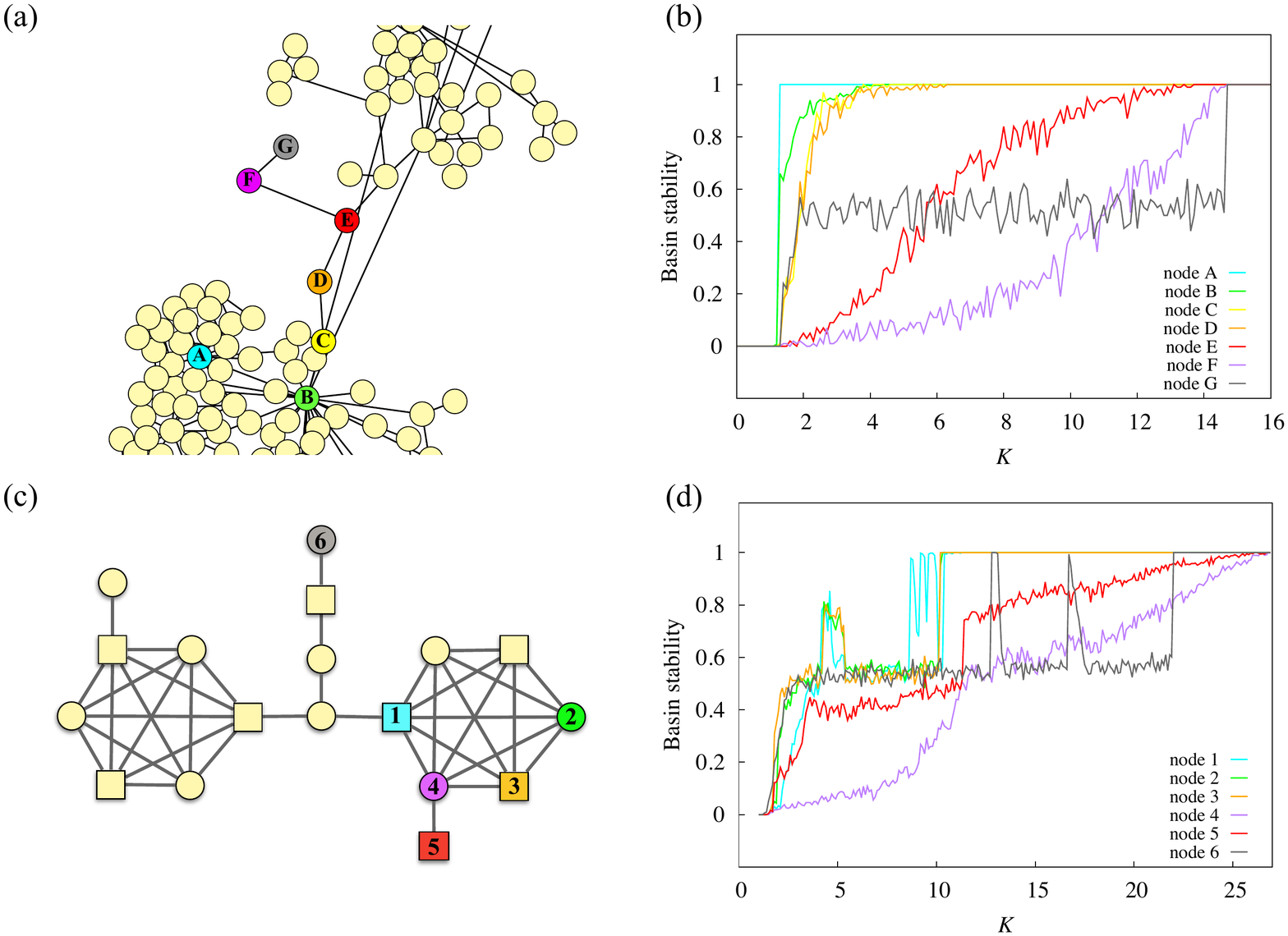}
\caption{Basin stability transition windows of some nodes in the Chilean power grid, (a) and (b), and the example network, (c) and (d). Same types of nodes in terms of basin stability transition patterns for the Chilean power grid and our example network, according to our judgment, are denoted with the same color, both for the nodes in (a) and (b), and the curves in (c) and (d). Numbers on the nodes are used for the identification in the corresponding basin stability transition plot, for each network separately.
}
\label{fig:model_network_result}
\end{figure}

In order to verify our results from the Chilean power grid that the basin stability transition is closely related to the community membership (figs.~\ref{fig:model_network_result}(a) and (b)) we construct a simple example network with prescribed community structures depicted in fig.~8(c). The example network consists of 18 nodes and most nodes are separated into two communities each of which has six fully connected nodes and a single attached dead-end node, except for four nodes branched from the bridge between the communities. The square nodes are assigned as consumers ($P_i = -1$ in Eq.~(\ref{eq:Kuramoto_type_equation})) and the circular nodes are producers ($P_i = 1$ in Eq.~(\ref{eq:Kuramoto_type_equation})). The example network symbolizes a network structure with two well-defined communities and outlier nodes on an interface branch. The interface branch represents a connected subgraph without clear community membership, just like the nodes F and G in the Chilean power grid (shown in fig.~\ref{fig:model_network_result}(a)). We measure basin stability transition window of the nodes in the example network and show representative cases in fig.~\ref{fig:model_network_result}(d). Most characteristics of the transition patterns observed from various nodes in the Chilean power-grid nodes (fig.~\ref{fig:model_network_result}(b)) are captured by some representative nodes in the example network (fig.~\ref{fig:model_network_result}(d)), e.g., concave versus convex transitions and the staircase pattern. 

To be more specific, the basin stability transition curve becomes convex or shows the staircase pattern, when the node is located far from a prescribed community and near to the branch from the bridge (thus with a weak community membership), in both Chilean power grid and our example network. Such outlier nodes seem to have a tendency of reaching a stable state (the basin stability equals to unity) at larger values of $K$ compared to the nodes with stronger community membership. In other words, our example network can be considered as the simplified version of the Chilean power grid at least in terms of the basin stability transition. In particular, we find some special transition patterns in both networks. Node G in the Chilean power grid (figs.~\ref{fig:model_network_result}(a) and (b)) shows a stability at a smaller value of $K$ compared to other nodes in the bridge. When a node is located near to the terminal node, the transition curve changes its shape form from concave down to concave up. However, the actual terminal node G in the Chilean power grid does not follow this trend and reaches medium level of basin stability at relatively small $K$ and maintains the plateau, until the basin stability suddenly reaches unity. Node 6 in our example network (fig.~\ref{fig:model_network_result}(d)) shows a similar pattern. Those nodes are commonly located at the terminus of the interface branch, and we speculate that this pattern is related to that the nodes cannot spread external perturbation from the neighbor nodes to the rest of the community. There are also some differences between the Chilean network and the example network, for example the spikes from the plateau to the maximum basin stability (fig.~\ref{fig:model_network_result}(d)).  For example, nodes 1 and 6 in the example network (figs.~\ref{fig:model_network_result}(c) and (d)) has maximum basin stability at $K \approx 5$ and $10$ (node 1) and $K \approx 13$ and $17$ (node 6), and they become unstable again. We find this phenomenon for the small example network too (not shown). Why this happens is beyond the scope of this paper, but not surprising in the light of the instabilities in synchronization on networks~\cite{Arenas2008}.

\section{Summary and discussions}
\label{sec:summary_and_discussions}

We have studied the basin stability transition throughout the coupling strength parameter space, focusing on the  basin stability transition window as a new metrics for characterizing the contribution of a node to the stability of a power grid. As previous works focus on the stability  for specific coupling strength values, our approach complements these. While a narrow transition window implies a sudden change in the stability, a wide transition window can provide an early signal of danger when the coupling strength is gradually weakened. By comparing the mesoscopic network property of community consistency with the transition window, we  found that the former is a good predictor of the latter (or vice versa), signifying the importance of community structures on the synchronization dynamics again. 

On a practical side, such community-consistency based predictions provide a proxy of the actual time-consuming simulations of calculating basin stability, especially for very large systems. Furthermore, once we assign the communities and the strength of nodes' memberships by community consistency, we are able to analyze the dynamics in the unit of such communities instead of the entire node set. We would also like to emphasize that the network example with given community structures and the bridge effectively captures the basin stability transition properties observed in the Chilean power grid.

Our approach can be improved by considering more realistic situations such as assigning different parameters of real fluctuations of power input and output, different edge weights based on admittance of the transmission line, etc. We see our work as a starting point, and anticipate more studies in the future. Finally, we would like to emphasize that our community-consistency measure is not limited to the power-grids, but could be used in all kinds of community detection problems.
 
\ack
We thank Beom Jun Kim, Mi Jin Lee, and Seung-Woo Son for fruitful discussions and comments. The authors were supported by Basic Science Research Program through the National Research Foundation of Korea (NRF) funded by the Ministry of Education (2013R1A1A2011947).

\end{document}